\documentclass[a4paper]{article}

\usepackage{icrc2013,xspace}
\usepackage{amsmath}
\usepackage{epstopdf}
\usepackage{enumitem}

\title{Influence of atmospheric aerosols on the performance of the MAGIC telescopes}

\shorttitle{Atmospheric Aerosols at the MAGIC Site}

\authors{
D. Garrido$^{1,2}$,
M. Gaug$^{1,2}$,
M. Doro$^{1,2,3}$,
Ll. Font$^{1,2}$,
A. L{\'o}pez-Oramas$^{4}$,
A. Moralejo$^{4}$
for the MAGIC Collaboration.
}

\afiliations{
$^1$ F{\'i}sica de les Radiacions, Departament de F{\'i}sica, Universitat Aut{\`o}noma de Barcelona, 08193 Bellaterra, Spain. \\
$^2$ CERES, Universitat Aut{\`o}noma de Barcelona-IEEC, 08193 Bellaterra, Spain. \\
$^3$ University and INFN Padova, via Marzolo 8, 35131 Padova (Italy) \\
$^4$ Institut de F{\'i}sica d'Altes Energies, 08193 Bellaterra, Spain.
}

\email{daniel.garrido@uab.cat}

\abstract{We investigate the performance of the MAGIC telescopes under three simulated atmospheric conditions: 
an increased  aerosol content in the lower part of the troposphere, the presence of thin aerosol over-densities at different
heights, and an extremely clean atmosphere. We show how the effective area of the telescope system is gradually 
reduced in the presence of varying concentrations of aerosols whereas the energy threshold rises. Clouds at different 
heights produce energy and altitude-dependent effects on the performance of the system.
}


\begin{document}
\maketitle

\section{Introduction} \label{sec.intro}

Imaging atmospheric Cherenkov telescopes (IACTs) use the atmosphere as calorimeters
to measure the characteristics of $\gamma$-ray-induced electromagnetic air showers. 
The mismatch between the true, real-time and simulated state of the atmosphere 
is currently the biggest source of systematic uncertainties~\cite{doro}.
While the effects of the (slowly-varying) molecular component on the performance of IACTs have been
precisely studied~\cite{konrad}, 
the aerosol component is complex to characterize, highly variable, 
and their effects on the reconstruction of air showers have not been fully taken into account so far by current analyses. 

\begin{figure}[h!t]
	\begin{center}
		\includegraphics[width=0.37\textwidth]{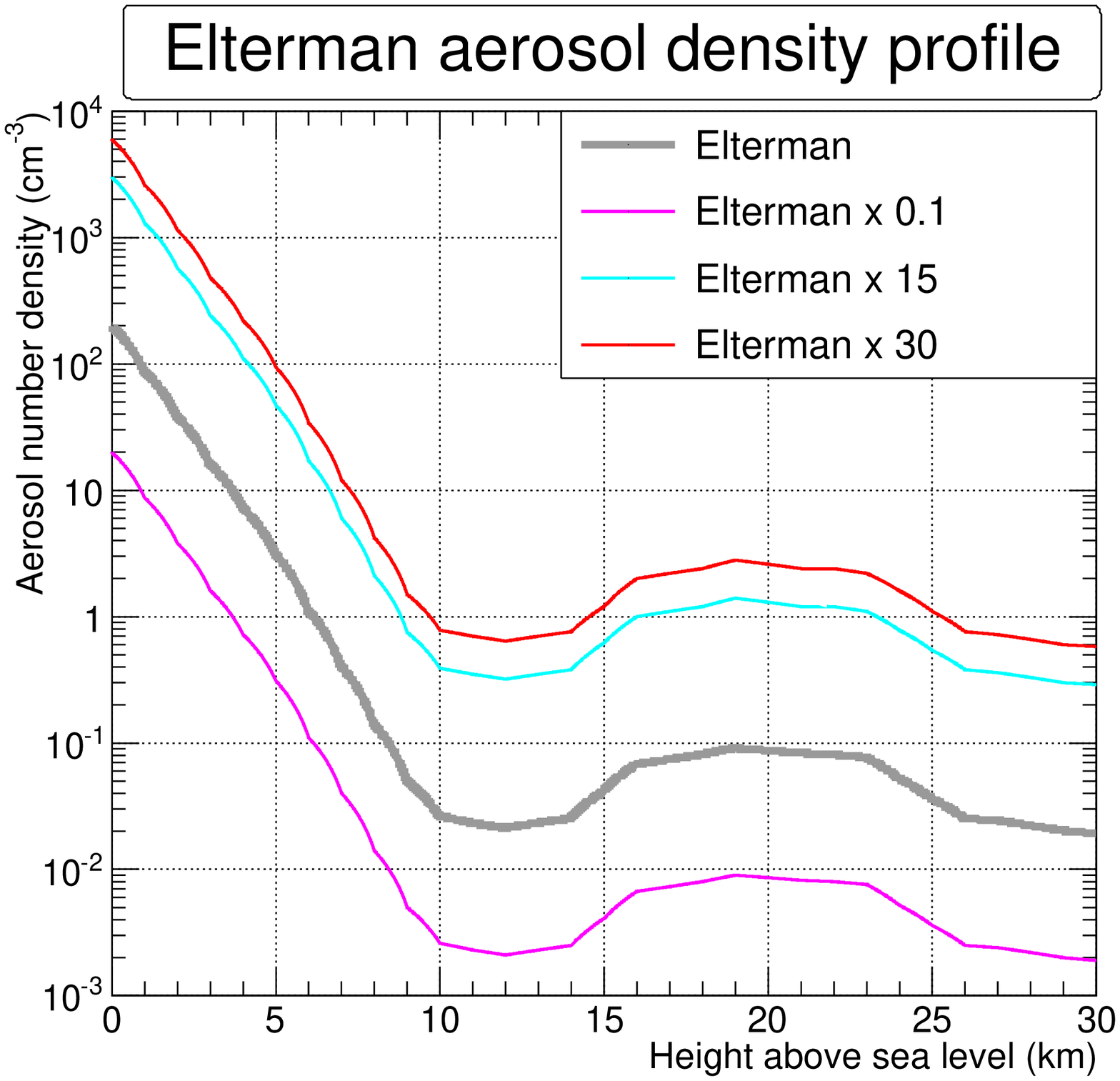} 
		\includegraphics[width=0.37\textwidth]{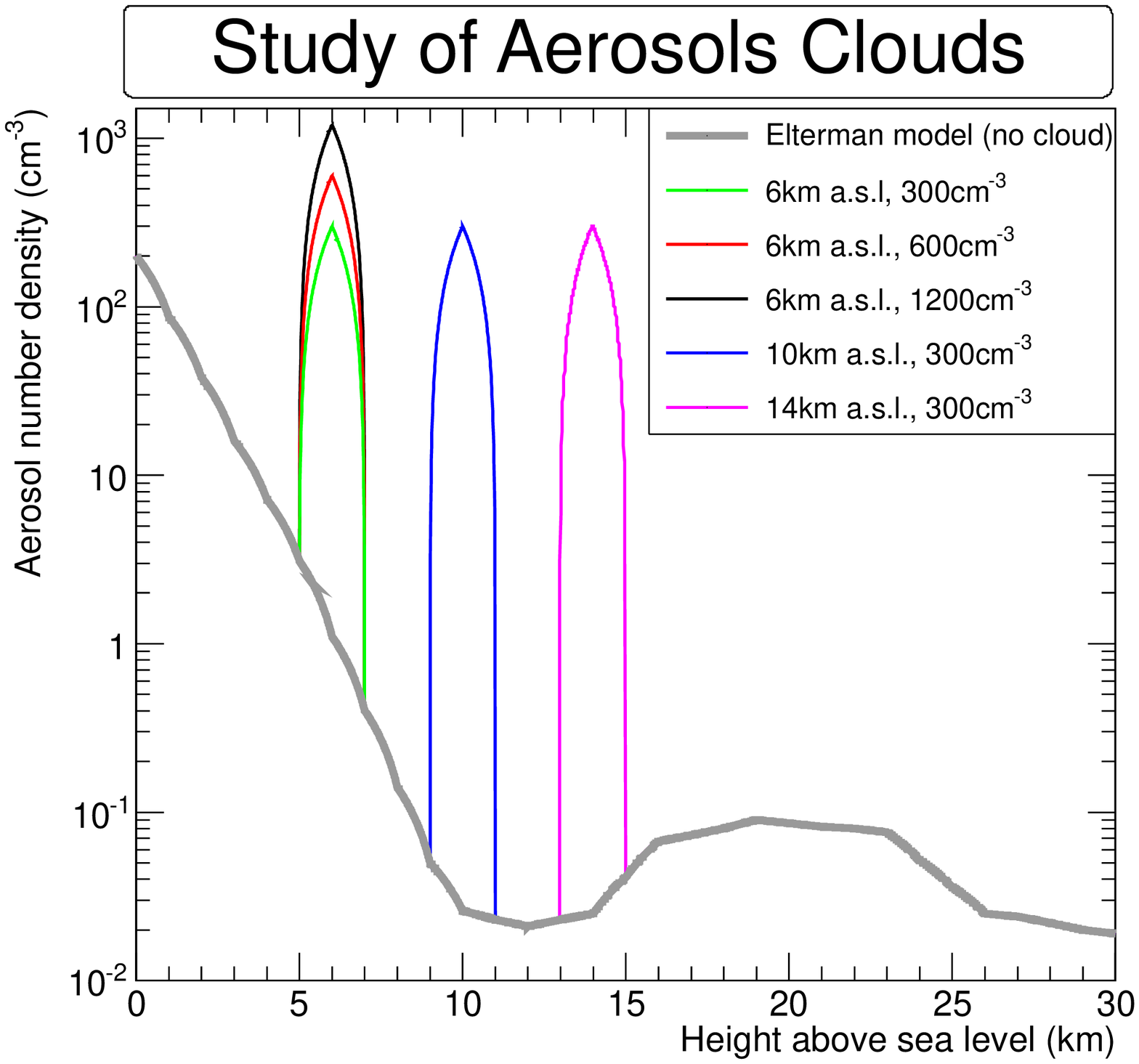}
	\end{center}
	\caption{\small{Particle number density as a function of height a.s.l. for the studied aerosol models.
			  	\emph{Top:} Elterman model (``Elt-1''), ``Elt-15'' (light blue), 
				``Elt-30'' (red) and ``Elt-0.1'' (pink). \emph{Bottom:} ``Elt-1'' (grey) compared to ``Cl-6'' or 
				``Cl-6/300'' (green), ``Cl-10'' (dark blue), ``Cl-14'' (pink) and ``Cl-6/600'' (red) and ``Cl-6/1200'' (black).
				The density was modified at just one altitude but the MAGIC atmosphere simulation software performs
				a linear interpolation between the previous and the following altitude bins.
				The curved shape is due to the logarithmic scale. \vspace{-0.4cm}
}}
	\label{fig.densityprofiles}
\end{figure}

The 
MAGIC telescopes consist of two 17~m diameter reflectors and are 
located at 2.2~km a.s.l. at the Roque de los Muchachos Observatory (ORM) on the Canary Island of La Palma~\cite{sitarek}. 
Data are analyzed using air shower simulations that take into account
only one single aerosol profile, namely the Elterman model~\cite{elterman}. 
It describes a low continental aerosol concentration of $200\text{ cm}^{-3}$ at sea level, 
followed by an exponential decay until the tropopause, above which it increases again to represent the stratospheric Junge layer. 
The density profile is shown in fig.~\ref{fig.densityprofiles} (\emph{top}, grey line). 
Several studies~(see e.g.~\cite{phd.rodriguez}) have shown however, that the 
atmosphere at the ORM is much cleaner than the Elterman model.

In order to understand the effects of aerosols on the data, we simulated a $\gamma$-ray flux
with a power-law differential energy spectrum of spectral index $-2.0$. 
In the simulation of the atmosphere, we then included different aerosol models. 
The data were processed in two energy reconstruction modes: 
one calculated from reconstruction tables which have been created from the very same atmosphere, 
and the other using tables obtained from a standard, clean atmosphere (which is hereafter referred to as ``wrong MC''). 
In this way, we could estimate the error made by analyzing data taken under non-optimal observation conditions, 
but not considering the correct aerosol distributions in the energy and flux reconstruction.

\section{Increasing the global aerosol density} 

We first produced 3 simple models, with the standard Elterman aerosol density 
profile (from now on ``Elt-1'') multiplied by a fixed factor, namely: 
15 for the ``Elt-15'' model, 30 for ``Elt-30'' and 0.1 for \mbox{``Elt-0.1''}. 
Both the \mbox{``Elt-15''} and the ``Elt-30'' models approximate Saharian dust intrusion events. 
While data taken under the first condition would not pass offline data
quality selection cuts, the ``Elt-30'' model applies to a situation where 
current MAGIC observation limits would even prevent data taking. 
The last model simulates an extremely clear atmosphere which matches better the observed aerosol concentrations at the ORM~\cite{lombardi}.

\begin{figure}[h!t]
	\begin{center}
		\includegraphics[width=0.4\textwidth]{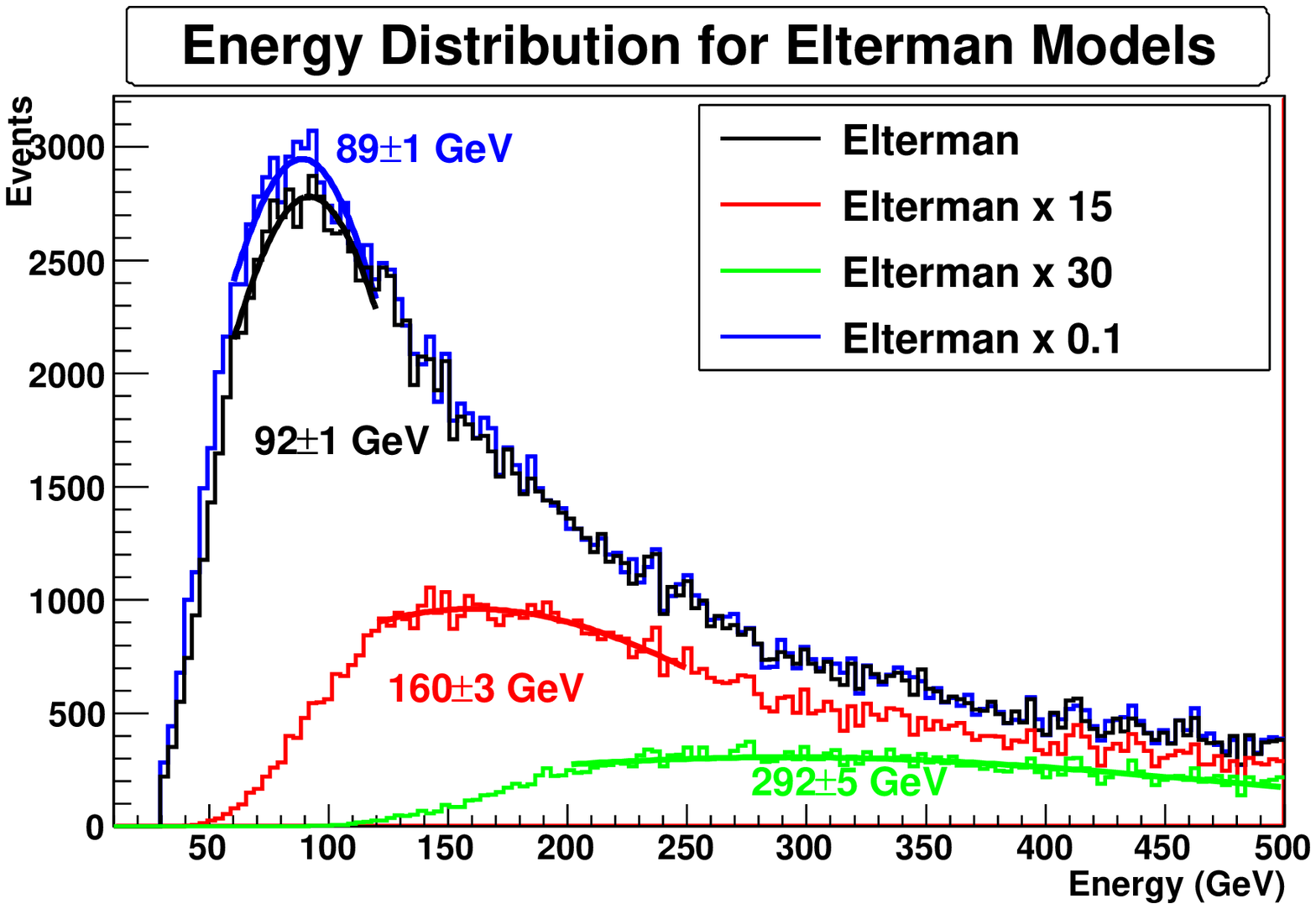}
		\includegraphics[width=0.4\textwidth]{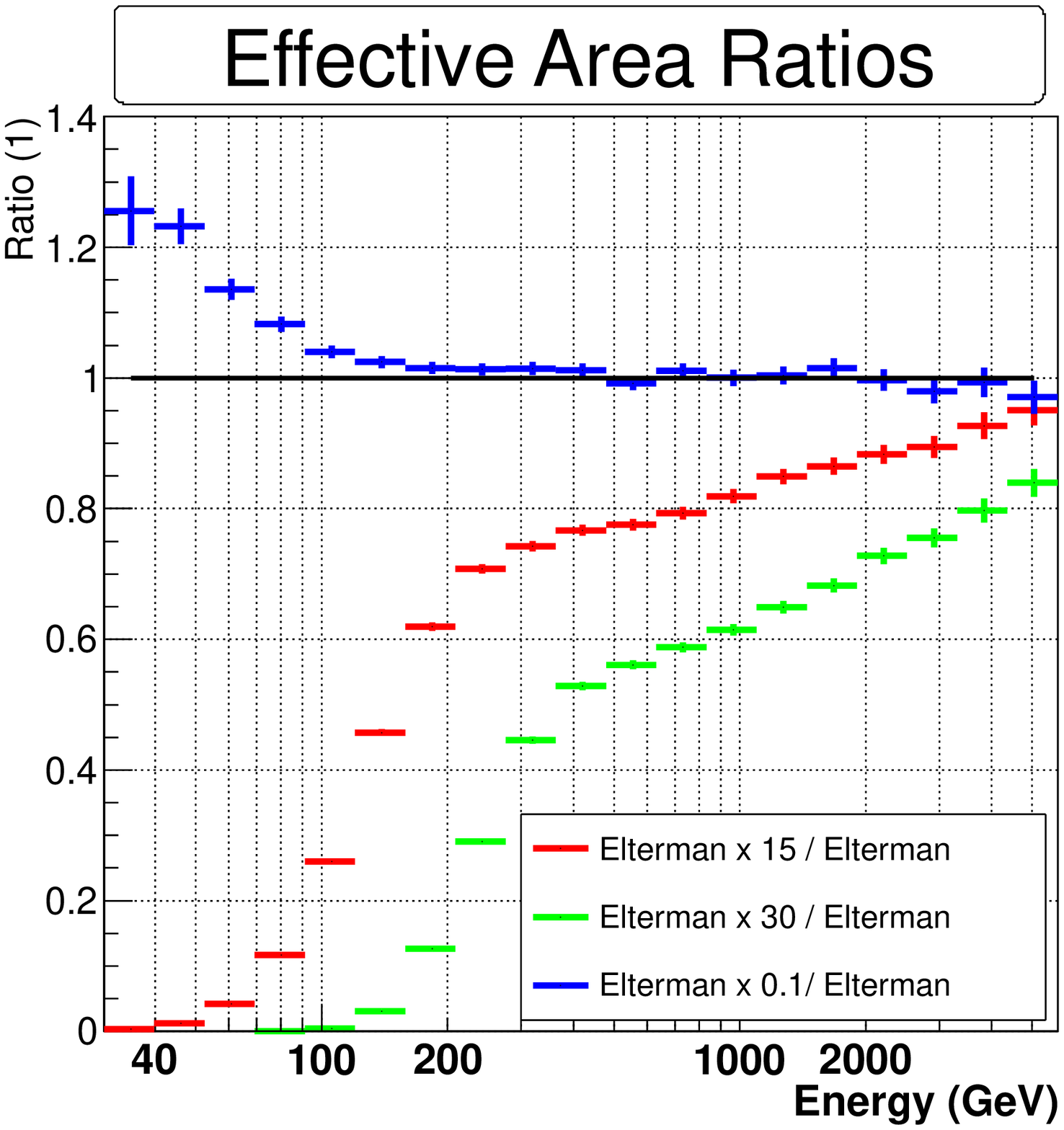}
	\end{center}
	\caption{ \small{Energy thresholds (\emph{top}) relative effective collection areas (\emph{bottom}) 
            for the ``Elt-1'', ``Elt-15'', ``Elt-30'' and ``Elt-0.1'' models. 	\label{fig.effarea_global}
            \vspace{-0.4cm}}}
\end{figure}

As expected, the energy threshold, defined as
the peak of the energy distribution (fig.~\ref{fig.effarea_global}, \emph{top}), increases with aerosol 
concentration, since the atmosphere becomes more opaque. For model ``Elt-30'', the telescopes are completely insensitive for primary 
$\gamma$-rays with energies below 150~GeV and still lose more than half of the events at energies around 300 GeV. 
A clear correlation between the energy threshold and integral atmospheric transmission at near-UV wavelengths has been found here~\cite{font}.


In fig.~\ref{fig.effarea_global} (\emph{bottom}) the ratios of effective areas of the 3 
models are shown, divided by the effective area of the \mbox{``Elt-1''} model, as a function of the $\gamma$-ray energy.
A significant decrease of the effective areas can be observed for the ``Elt-15'' and ``Elt-30'' cases 
at all energies, especially at those below the threshold, as well as an increase  at the very lowest energies of up to 30\,\% for the 
cleaner \mbox{``Elt-0.1''} model.

We can also observe that the ratios are not constant, even far above the energy threshold: 
at low energies, the effective area ratio gets strongly reduced until a \emph{break energy}, $E_{br}$. 
Above it, the effective area ratios keep increasing with energy, however at a smaller pace. 
The retrieved values of $E_{br}$ for all models can be found in tab.~\ref{tab.fits}. 

\begin{figure}[h!t]
	\centering
	\includegraphics[width=0.49\textwidth]{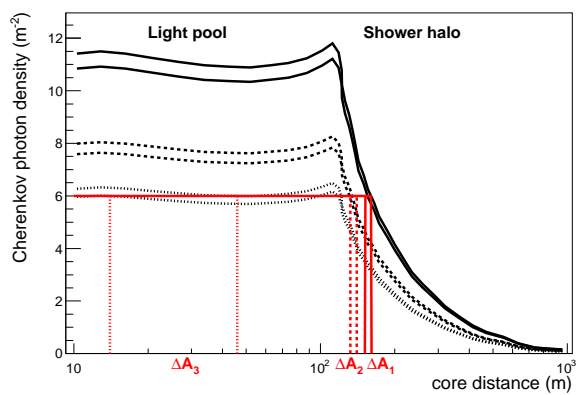}
	\caption{\small{Sketch of the Cherenkov photon density of a $\gamma$-ray induced air shower as a function
			of the core distance. Pairs of solid, dashed and dotted lines show 3 events with 
			decreasing energies. In each pair, the upper and lower lines correspond
			to a clean and hazy atmosphere, respectively. The horizontal red line represents 
			the telescopes trigger threshold. The regions labeled $\Delta A_1$ and $\Delta A_2$ 
                        present possible changes of trigger area, 
                        due to the presence of an increased atmospheric aerosol content. 
                        The last change labelled $\Delta A_3$ only insinuates the strong reduction of effective area, 
                        since here it is completely dominated by shower fluctuations observed throughout the entire light pool.
                        \label{fig.chdensity}	 \vspace{-0.4cm}
}}
\end{figure}

To illustrate how a break in the effective area ratios can be obtained, 
fig.~\ref{fig.chdensity} shows the typical Cherenkov photon density for a vertically incident $\gamma$-ray shower, 
 as a function of the core distance~\cite{konrad}. Two standard regions can be distinguished: 
the so-called \emph{Cherenkov light pool} (the plateau region before the peak) and the \emph{Cherenkov halo}, 
which quickly decays after the peak. 
An air shower of small energy, like the one represented by the dotted line, at the limit of the 
telescope sensitivity (red line) will only trigger the telescope readout if the telescope happens to be 
found at a small distance from the core, or an upward fluctuation in flight yield has occurred. 
An aerosol over-density will strongly reduce the collection area for these energies, since the resulting photon density 
\emph{of the light pool} falls below the trigger threshold. 
When the energy of the event is big enough (dashed and solid lines in the figure), 
an event inside the light pool will always be detected, however an increase of the aerosol content of the atmosphere will reduce 
the region \emph{of the halo} which can still trigger the telescopes. This leads to 
the smaller slope in fig.~\ref{fig.effarea_global} (\emph{bottom}), after $E_{br}$. 

\begin{figure}[h!t]
\centering
\includegraphics[width=0.48\textwidth]{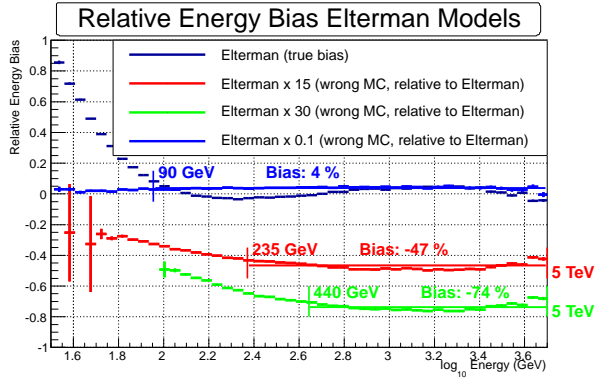}
\caption{\small{Energy biases for the different aerosol models. The standard bias is shown in black, 
while the other cases show the obtained biases w.r.t. 
to that standard case. The energy has been reconstructed 
(wrongly) assuming standard atmospheric conditions. The flat part of the biases has 
been fitted to a straight line, and the fit result and fit limits are displayed in colors.\label{fig.bias} 
\vspace{-0.1cm}
}}
\end{figure}

The presence of ground-layer aerosols causes also a bias in the energy reconstruction (see fig.~\ref{fig.bias}), 
hence effective areas are (wrongly) evaluated at lower reconstructed energies. 
See~\cite{dgarrido,fruck} for further details.

\begin{figure}[h!t]
	\begin{center}
		\includegraphics[width=0.48\textwidth]{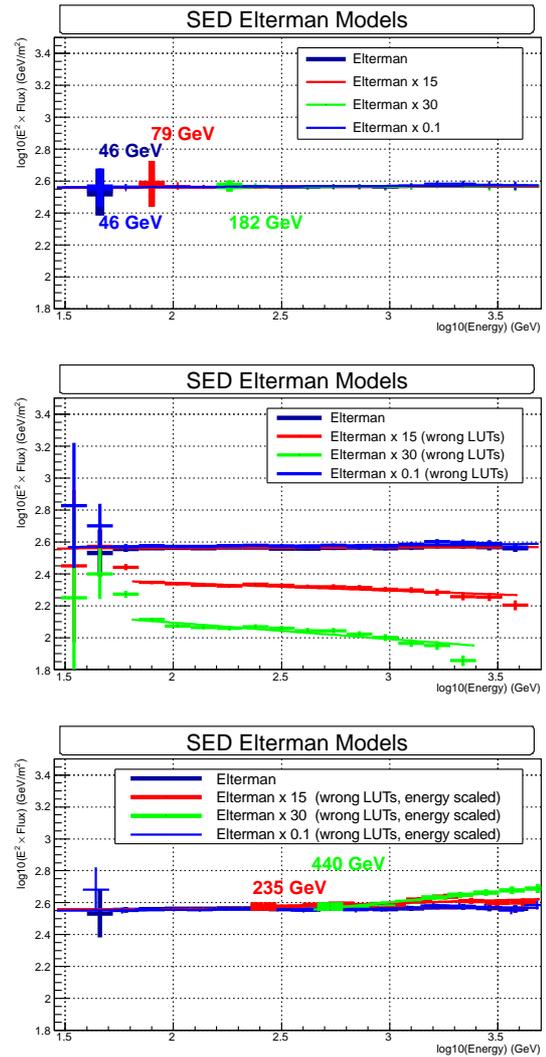}
	\end{center}
	\caption{\small{Spectral energy density reconstruction for the Elt-1, Elt-15, Elt-30 and Elt-0.1 
				models using adapted MC for the analysis (\emph{top}), using standard MC
				(\emph{center}) and 	using standard MC with a constant energy and effective 
				area correction factor (\emph{bottom}). The smallest reconstructed energy bin 
				is shown enhanced, with the corresponding energy beside.  The main 
				parameters of these spectra can be found at tab.~\ref{tab.fits}.
	\label{fig.flux123}
\vspace{-0.5cm}}}
\end{figure}

These effects combine in the estimation 
of spectra and SED, summarized in tab.~\ref{tab.fits} and shown in fig.~\ref{fig.flux123}.
We can see that, if standard MC is used, the spectral 
index increases significantly (the spectra steepens), and the flux of the source is systematically 
under-estimated (fig.~\ref{fig.flux123} (center)).

If the reconstructed energies are shifted equally by a same correction factor, derived from fits to the energy bias in fig.~\ref{fig.bias}, 
and the effective area is corrected by another constant correction factor, the original SED can be recovered
(fig.~\ref{fig.flux123}, \emph{bottom}) at the price of an unavoidable increase of the energy threshold. 

\section{Aerosol layers height and density}

Next, we introduced 3 models containing an 
aerosol layer 
of $300\text{~particles/cm}^3$ at 3 different heights. Their density 
profiles are shown in fig.~\ref{fig.densityprofiles} (\emph{bottom}): a layer at 6~km (model ``Cl-6''), 10~km (``Cl-10'') and 14~km (``Cl-14''). 
Since Cherenkov light from air showers is produced at different heights and travels several kilometers through the atmosphere, height-dependent effects
are expected. Stars, in turn, would appear  equally under all 3 conditions, about half a magnitude dimmer.

\begin{figure}[h!t]
	\begin{center}
		\includegraphics[width=0.8\linewidth]{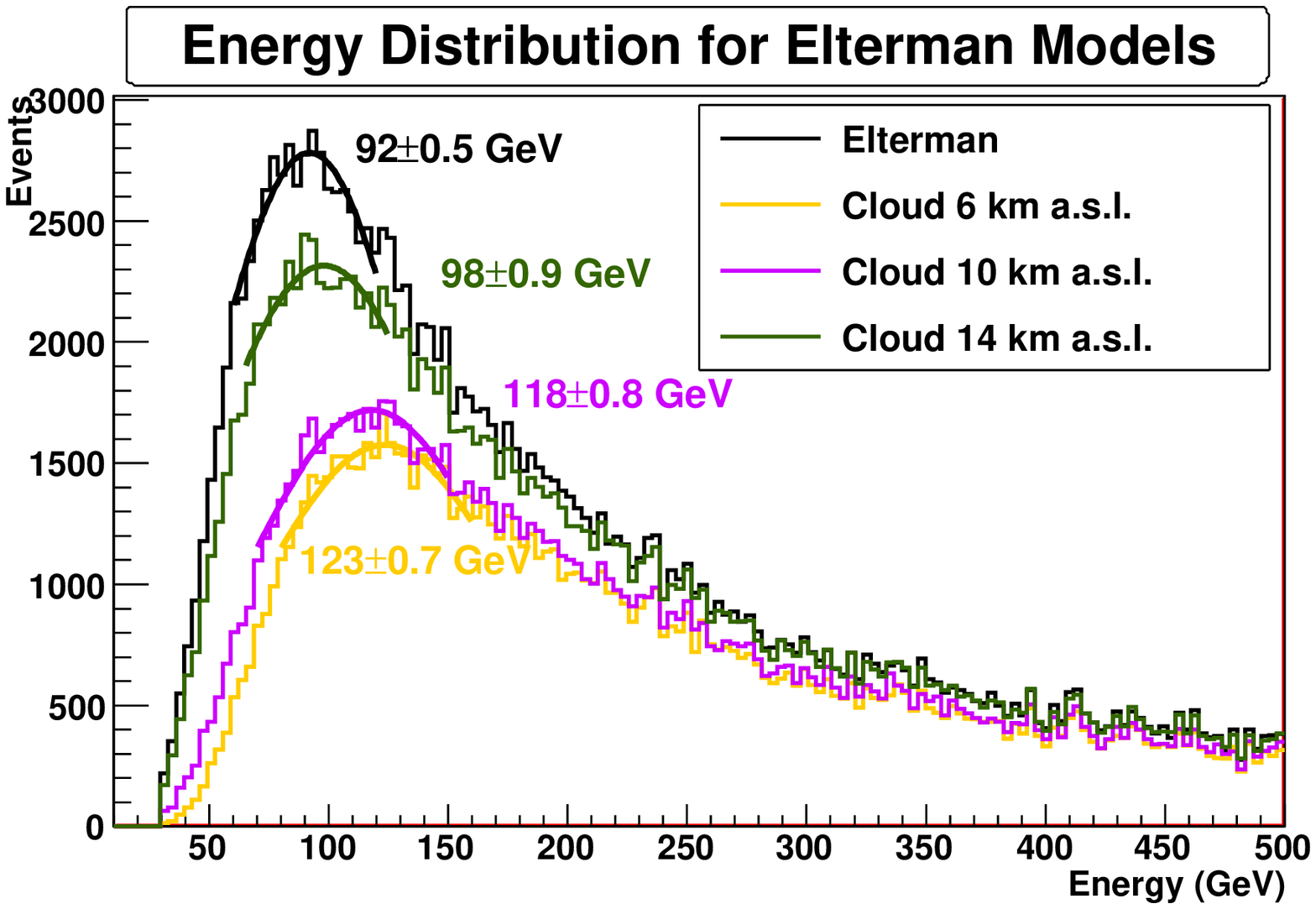}
		\includegraphics[width=0.8\linewidth]{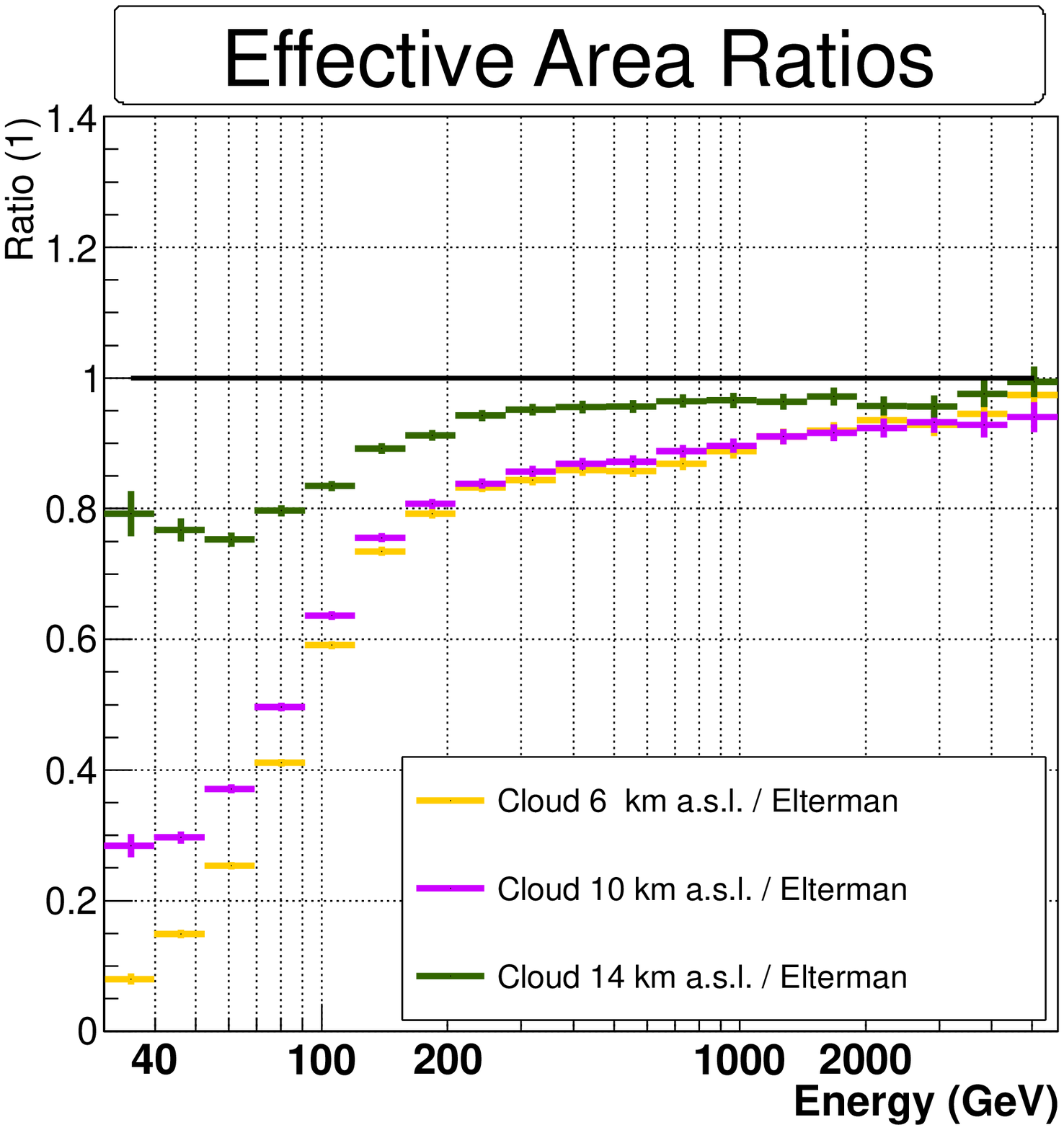}
	\end{center}
	\caption{\small{Energy thresholds (\emph{top}) and effective area ratios (\emph{bottom}) for 
	 models ``Cl-6'', ``Cl-10'' and ``Cl-14'' w.r.t. the standard atmosphere.
	\label{fig.effarea_cloud1}
\vspace{-0.1cm}}} 
\end{figure}

\begin{figure}[h!t]
\centering
\includegraphics[width=0.48\textwidth]{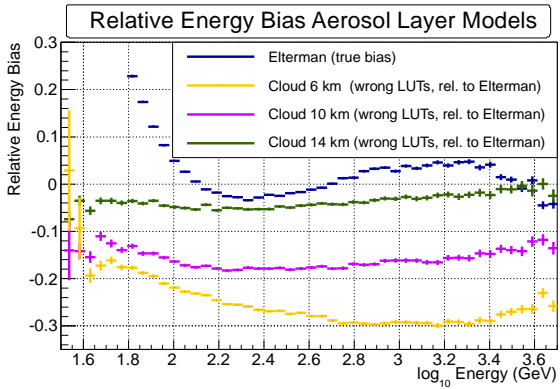}
\caption{\small{Energy biases for the different aerosol layer models. The standard bias is shown in black, 
while the other cases show the obtained biases w.r.t. 
to that standard case. The energy has been reconstructed 
(wrongly) assuming standard atmospheric conditions. \label{fig.layerbias} 
\vspace{-0.4cm}
}}
\end{figure}

\begin{table}[h!]
	\centering
	\begin{tabular}{l || c c c | c | c}
	
\hline\hline  
Model                                 & $D_{300}$ & $\alpha$     & $\Delta E_{low}$  & $E_{th}$   & $E_{br}$       \\
\hline\hline	
Elt-1                                   & $2.40$         & $2.00$        & 46                        & 92.0         & -                 \\
                                          &                     &                    &                              & $\pm0.5$  &                   \\
\hline
Elt-0.1$^{\star}$                 & $2.38$        &  $2.00$        & 46                     & 88.9          & $103 $    \\ 
Elt-0.1$^{\dagger}$            & $2.40$        &  $2.00$        & 44                     & $\pm 0.5$ & $\pm 29$  \\
\hline
Elt-15$^{\star}$                  & $\bf{2.64}$  & $2.05$        &  -                       & 160           & $220.0 $         \\
Elt-15$^{\dagger}$             & $2.39$        & $\bf{1.96}$  & 235 (83)           &  $\pm 3$   & $\pm 6.2 $    \\
\hline
Elt-30$^{\star}$                  & $\bf{2.92} $ &  $\bf{2.09}$ &  -                     & 292           &  $453 $           \\
Elt-30$^{\dagger}$             & $2.44$        &  $\bf{1.86}$ & 440 (198)         &  $\pm 5$   & $\pm 15$     \\
\hline\hline             
Cl-6$^{\star}$                     & $\bf{2.54}$  & $\bf{2.05}$  &  -                    & 123.4        & $156.3$      \\
Cl-6$^{\dagger}$                & $2.39$        & $1.97$         & 150  (83)        & $\pm 0.7$ & $\pm 5.3$   \\
\hline                
Cl-10$^{\star}$                   & $\bf{2.53}$  & $1.98$        &  -                     & 118.0        & $157.9$      \\
Cl-10$^{\dagger}$              & $\bf{2.45}$  & $\bf{1.91}$  &100  (83)         & $\pm 0.8$&  $\pm 8.9$   \\
\hline
Cl-14$^{\star}$                   & $\bf{2.44}$  & $1.99$        &  -                     & 98.0          & $232$         \\
Cl-14$^{\dagger}$              & $2.41$        & $1.98$        & 80  (62)           & $\pm 0.9$ & $ \pm 53$   \\
\hline\hline
Cl-6/600$^{\star}$              & $\bf{2.69}$  & $\bf{2.06}$  &  -                    & 170            & $232.1$      \\
Cl-6/600$^{\dagger}$         & $2.40$        & $\bf{1.97}$  & 250 (111)        &  $\pm 1$    & $\pm 6.6$   \\
\hline                     
Cl-6/1200$^{\star}$            & $\bf{3.07}$  & $\bf{2.15}$  &    -                  & 305            & $ 343$        \\
Cl-6/1200$^{\dagger}$       & $\bf{2.47}$  & $\bf{1.93}$  & 350 (198)        &  $\pm 3$    & $ \pm 26$   \\
\hline\hline
	
	\end{tabular}
	\caption{\small{Fit results of the simulated aerosol models (``Cl-6'' is the same as ``Cl-6/300''). Spectra $= \frac{dN}{dEdA}=10^{-D_{300}}
				\left(\frac{E}{300\,GeV}\right)^{-\alpha}\text{ GeV}^{-1}\text{m}^{-2}$;
				$\Delta E_{low}$: lowest reconstructed energy bin of the spectra when standard (adapted) MC was used,
				 in units of GeV ; $E_{th}$, energy threshold in units of GeV; $E_{br}$, effective area ratios break energy, in units of GeV; 
				a star ($^{\star}$) refers to wrong (clean atmosphere) energy reconstruction tables; a dagger ($^{\dagger}$) refers to wrong 
                                energy reconstruction tables, plus posterior energy scaling. The fits labeled with $^{\dagger}$ 
				have been obtained using a lower energy limit to avoid that the artificial high-energy cut-off modified 
				the estimated spectral slope. The bold values show deviations bigger than 4$\,\sigma$. The values of 
				$E_{th}$ and $E_{br}$ are unique for each aerosol model, independent of the event reconstruction method.
				Spectral index and flux at 300~GeV were always correctly recovered when we used adapted MC so they
				are not shown here. See also~\cite{font}. \vspace{-0.3cm}}}
	\label{tab.fits}
\end{table}

The energy thresholds, $E_{th}$, for these models are shown in fig.~\ref{fig.effarea_cloud1} (top) and listed in tab.~\ref{tab.fits}. 
They increase as the height of the layer decreases, as expected, since low-energy showers develop further up in the atmosphere.
However, only a small difference is seen between the layer at 6~km and 10~km altitude.
Similarly, the effective area is reduced equally by the low- and mid-altitude layer, except for energies below the threshold.
A break energy is visible, as found in the previous study. However, the high layer seems to affect collection area ratios
 differently: rather two smoothly joint levels from the minimum to the threshold, and later from about 3 times the threshold to the high energies, 
are observed. The last points showing values higher than unity, are due to image leakage cuts in the analysis. 

When trying to reconstruct the source spectra, we found that the low-altitude layer behaves like the models with increased overall aerosol density: 
If the energy is scaled according to the found bias (not shown here, refer to~\cite{font}), the source spectrum can be recovered with a generic 
(small) overall correction factor on the effective area (cf. tab.~\ref{tab.fits}). The method fails for the mid-altitude layer, for which a different 
algorithm has been developed~\cite{fruck}. As the high layer affects fluxes only hardly, especially above the threshold, corrections are small, 
and statistics have not been sufficient to decide whether the simple scaling of the energy introduces a systematic error bigger than the statistical one.



In a last study, we investigated the effects of 3 aerosol layers located at the same height of 6~km a.s.l., 
with different aerosol number densities (fig.~\ref{fig.densityprofiles} bottom).  
The first two models correspond to scarcely and moderately opaque atmospheric 
conditions, under which data taking can go on, however data will be gradually discarded by data quality cuts. 
The third model 
describes an atmospheric situation under which telescope operators stop taking data. 
The retrieved effective area ratios show a similar behavior to the ones found for the scaled Elterman 
models.
The 
spectra could again be recovered with both the use of correct energy tables and the global energy scaling, at the 
price of higher energy thresholds (see tab.~\ref{tab.fits}). 

%


\section{Discussion and Conclusions}


Three cases of aerosol intrusions can be distinguished at the ORM: Saharian dust intrusions, clouds or layers of haze 
at different altitudes, and seldom occurring layers of volcanic debris, ejected into the lower stratosphere. 
We found that the first case, i.e. aerosol enhanceemnts in the ground layer, and aerosol layers until about 6~km a.s.l., have similar effects on the recorded shower images (see also~\cite{font}): 
Analysis thresholds increase linearly with integral atmospheric 
transmission, and reconstructed energies show a similar negative bias, unless the aerosol layer is included in the energy reconstruction tables. 
Effective areas get reduced w.r.t. the clear atmosphere, the more the lower the shower energy. The relative reduction 
is stronger below a certain break energy, found always higher than the threshold. 
Above it, effective areas gradually approach the clear atmosphere case at the highest energies.
This behavior can be understood well by the transition from triggering exclusively inside the Cherenkov light pool to becoming sensitive 
to the shower halo. Unless corrected, a small, gradual steepening of a power-law spectral index with atmospheric transmission can be observed.



Intermediate layers around 10~km a.s.l. produce energy-dependent effects and hence introduce 
some steepening of the spectra, if not simulated correctly. 
An event-wise correction method is required here, presented in~\cite{fruck}, based on real-time monitoring
of the height-resolved atmospheric transmission~\cite{fruck,doro}.

Finally, we should mention that although this study was performed using specific simulation
of the MAGIC telescopes, its conclusions can be extended to other IACTs and  
the next generation of ground-based $\gamma$-ray facilities like the Cherenkov 
Telescope Array (CTA)~\cite{ctaconcept}.

\bibliographystyle{plain}

\begin{thebibliography}{99}
\setlength{\itemsep}{2.0pt plus 1.0pt minus 2.0pt}
\bibitem{doro} M.\,Doro et al., these proceedings ID 151.
\bibitem{konrad} K.\,Bernl\"ohr, Astrop. Phys., 12(4):255--268, 2000.  
\bibitem{sitarek} J.\,Sitarek et al., these proceedings ID 74
\bibitem{elterman} L.\,Elterman, Applied Optics 3 (1964) 745.
\bibitem{phd.rodriguez} A.M.\,D\'iaz, PhD. Thesis, 2006, Universidad de La Laguna, (Spain).
\bibitem{font} Ll.\,Font, et al., these proceedings, ID 90.
\bibitem{lombardi} G.\,Lombardi, V.\,Zitelli, S.\,Ortolani, M.\,Pedani, A.\,Ghedina, A\&A 483:651--659, 2008.
\bibitem{dgarrido} D.\,Garrido, Master thesis, 2011, Universitat Aut\`onoma de Barcelona (Spain).
\bibitem{fruck} C.\,Fruck, et al., these proceedings, ID 1054.
\bibitem{ctaconcept} {Acharya, B.S.} et~al., Astrop. Phys., 43:3 -- 18, 2013.

\end{thebibliography}

\end{document}